\documentclass[12pt]{article}
\usepackage{graphicx}
\usepackage{amssymb}
\usepackage[square, comma, sort&compress]{natbib}

\input epsf.sty

\newcommand{\bd}{\begin{displaymath}}
\newcommand{\ed}{\end{displaymath}}
\newcommand{\ba}{\begin{eqnarray}}
\newcommand{\ea}{\end{eqnarray}}
\newcommand{\be}{\begin{equation}}
\newcommand{\ee}{\end{equation}}
\newcommand{\ben}{\begin{eqnarray}}
\newcommand{\een}{\end{eqnarray}}

\newcommand{\p}{\partial}
\newcommand{\sectiono}[1]{\section{#1}}
\def\sqr#1#2{{\vcenter{\vbox{\hrule height.#2pt
         \hbox{\vrule width.#2pt height#1pt \kern#1pt
            \vrule width.#2pt}
         \hrule height.#2pt}}}}

\begin{document}

\begin{center}
\large{\bf Cosmological imprints of pre-inflationary particles}

\vspace{10mm}

\normalsize{Anastasia Fialkov\footnote{anastasia.fialkov@gmail.com}, Nissan Itzhaki\footnote{nitzhaki@post.tau.ac.il} and Ely D.
Kovetz\footnote{elykovetz@gmail.com}\\\vspace{2mm}{\em Tel-Aviv University, Ramat-Aviv, 69978,
Israel}}

\end{center}

\vspace{10mm}

\begin{abstract}

We study some of the cosmological imprints of pre-inflationary particles. We show that each such particle provides a seed for a spherically symmetric cosmic defect. The profile of this cosmic defect is fixed and its magnitude is linear in a single parameter that is determined by the mass of the pre-inflationary particle. We study the CMB and peculiar velocity imprints of this cosmic defect and suggest that it could explain some of the large scale cosmological anomalies.

\end{abstract}

\newpage

\baselineskip=18pt

\newpage
\sectiono{Introduction}
\bigskip

In this paper we study the cosmological imprints of a pre-inflationary particle (PIP).  By a PIP we mean a particle that existed at the latest during the end of the pre-inflationary period and the beginning of inflation. The reason we focus on PIPs is that this is the simplest and most generic prediction concerning the speculative pre-inflationary era, and indeed such particles are created in many pre-inflationary scenarios.

At first sight there seems to be little motivation for this study. The reason is that
one of the most basic features of cosmic inflation is that it erases all details of the pre-inflationary period. It is this feature that makes inflation such a predictive model. The parameter that determines how precise this statement is, is  $X=e^{-\Delta N}$, where $\Delta N = N_{tot}-N_{BB}$ is the difference between the total number of e-foldings accumulated during inflation and the number of e-foldings required to resolve the big-bang puzzles (horizon and flatness problems, etc). Roughly speaking, the amount of pre-inflationary imprints in the visible universe scales like a polynomial in $X$, the details of which depend on the pre-inflationary period. For example, imprints of PIPs in the visible universe scale as $X^3$ (since they get diluted like $1/a^3(t)$).

In large field models of inflation $\Delta N$ is typically large and the probability of finding pre-inflationary imprints in the visible universe is negligible. However, in small field models of inflation $\Delta N$ can be fairly small and there might be pre-inflationary imprints in the visible universe.

Pre-inflationary imprints, if they exist, are expected to show up in the largest scales in the visible universe. Interestingly enough, some observations at the largest scales  appear to disagree with   $\Lambda$CDM \cite{Bennett:1996ce,Bennett:2003bz,Spergel:2003cb,Copi:2006,Copi:2008hw,Watkins,Lavaux:2008th,OC,Vielva,Eriksen04}.  The results are far from being conclusive (partially because of cosmic variance). Still, we find them intriguing enough to entertain the possibility that $\Delta N$ is small, and attribute at least some of them to pre-inflationary effects in general and PIPs in particular. In fact, the most generic prediction of a small $\Delta N$ is a significant drop in the power spectrum at the largest scales \cite{Contaldi:2003zv} in accord with COBE  \cite{Bennett:1996ce} and WMAP \cite{Bennett:2003bz,Spergel:2003cb} (for a recent analysis see \cite{Copi:2008hw}).

In section 2 we show how a PIP affects the evolution of the inflaton during inflation. We show that to leading order this effect is not sensitive to the details of the pre-inflationary period, but only to a single parameter, $\lambda$, determined by the mass of the PIP. In the linearized  approximation the dependence is proportional to $\lambda$, which simplifies the calculations throughout the paper.
We also calculate the ideal signal to noise ratio ($S/N$) associated with such a particle. By ideal $S/N$ we mean a situation in which we have access to the full cosmological data in the visible universe.

In section 3 we focus on the PIP's imprints on large scale structure. We show that a PIP provides a seed for a spherically symmetric cosmic defect (SSCD), in the form of an extended region in space with higher or lower mean density. The shape of the SSCD is fixed and its magnitude is linear in $\lambda$. The sign of $\lambda$ determines whether this is an over dense or an under dense region. We study in detail the gravitational potential and the profile of the SSCD.
In section 4 we study the PIP's imprints on the CMB and investigate their dependence on the location of the SSCD. We show that the main effect lies in the low-$l$ modes and that there is a competition between the Sachs-Wolfe (SW) and the Integrated SW (ISW) effects induced by the SSCD. We calculate the CMB signal to noise ratio as a function of the location of the SSCD and find that the SW-ISW competition leads to some interesting effects that could help in identifying the SSCD's imprints in the CMB sky. We also show that a SSCD located at a particular distance from us provides a possible explanation to the WMAP cold spot.

The peculiar velocity imprints of the SSCD are calculated in section 5. We find that the peculiar velocity signal of a relatively nearby SSCD is larger than its CMB signal. In section 6 we consider the possibility that the anomalous peculiar velocity reported recently by \cite{Watkins,Lavaux:2008th} is induced by a SSCD. We calculate the CMB $S/N$ associated with a SSCD assuming it is responsible for this anomalously large peculiar velocity as a function of its location. We find that typically such a SSCD is CMB detectable, when focusing on the right observables.
We conclude in section 7.

\sectiono{Point-like particle during inflation}

In this section we generalize the discussion of \cite{Sunny} and  study the effects a single Pre-Inflationary Particle (PIP) has on the evolution of scalar inhomogeneities.

For simplicity we assume that the  PIP is located at the origin of a spatially flat, expanding universe
\be
ds^2 = -dt^2+a^2(t)\delta_{ij}dx^idx^j.
\ee
To calculate the scalar inhomogeneities we find it most convenient to use the spatially flat slicing gauge in which the inhomogeneities are parameterized by
\be
\phi(x,t) = \bar{\phi}(t)+\delta \phi(x,t),~~~~\mbox{and}~~~~\delta
h_{ij}=0.
\ee
In this gauge the equation of motion for the perturbation in the inflaton in the presence of a PIP (see Appendix A.1 for the derivation) reads
\be \label{eqmotion}
\delta\ddot\phi+3H\delta\dot\phi-\frac{1}{a(t)^2}\nabla^2\delta\phi
 = -\lambda \frac{\delta^3(x)}{a(t)^3} ,
\ee
where
\be\label{wq}
\lambda = m'-\frac{1}{2}
\frac{ V'}{V}m,
\ee
is a dimensionless  effective parameter controlling the influence of the pre-inflationary particle. We work in units where $c=(8\pi G)^{-1/2}=1$ and $'$ means derivative with respect to $\phi$.

Expanding in the slow-roll parameters, the leading contribution to $\lambda$ comes from $m'$. The reason is simple: if the mass depends on the inflaton then there is a direct coupling between the PIP and the inflaton, while constant mass means that the interaction is indirect, via gravity.

Eq. (\ref{eqmotion}) differs from the standard equation of motion for the perturbation $\delta\phi$ by the addition of a source term. Being linear,  its solution is a sum of a homogeneous solution and a particular inhomogeneous solution.
The homogenous solution of the above equation describes the quantum fluctuations of the inflaton field. These fluctuations  have a gaussian distribution function with a vanishing one-point function and a non-vanishing two point function (power spectrum)\footnote{The normalization factor in this equation depends on the normalization convention of the Fourier integral. We work with $F(\vec x) = \int \frac{d^3k}{(2\pi)^{3/2}} e^{i\vec k \vec x}F(\vec k)$.}
\be\label{power}
\langle \delta\phi_{\vec k} \delta\phi_{\vec k'}\rangle=\left.\frac{H^2}{2 k^3} \right| _{k=a(t) H} \delta (\vec k-\vec k'),
\ee
where $H$ is the Hubble constant during inflation. It is this  power spectrum that is believed to provide the seed  for most, if not all, the large scale structure in the universe.

Imposing the natural initial condition for the inhomogeneous solution (i.e. that it scales like $1/r a(t)$  in the limit $ra(t)\ll 1/H$) we find that the late time  inhomogeneous solution induced by the PIP reads (for details see \cite{Sunny})
\be\label{PIP}
\langle \delta\phi_{PIP} (k)\rangle =-\left.\frac{\lambda}{k^3}\left(\frac{H
}{{\sqrt{32\pi}}}\right)\right| _{k=a(t) H}.
\ee
As usual the equation is evaluated at horizon crossing. Note that  $\lambda$ could depend on $k$ as well.
Notice further that, unlike in the homogeneous solution, here we are talking about a one-point function and not a two-point function. The non-vanishing one-point function depends only on $k \equiv\left|\vec k\right|$. Thus, the PIP creates a spherically symmetric defect in the inflaton field.

Eqs. (\ref{power}) and (\ref{PIP}) imply that both $\delta\phi_{PIP}$ and $\delta \phi$ scale like $H$. Therefore, for the PIP's imprints to be noticeable we need the dimensionless parameter $\lambda$ to be  large, independently of $H$. To make this statement more precise we compute the signal to noise ratio ($S/N$) in an ideal situation, in which a PIP is located at the origin of the survey and we have full access to comoving  modes in the range $k_1<k<k_2$. By signal here we mean
$\delta\phi_{PIP}$ and by noise the quantum fluctuation power spectrum.
We have
\be\label{ideal-stn}
\left(\frac{S}{N}\right)^2_{ideal}=4\pi \int dk k^2 \frac{\delta\phi^2_{PIP}(k)}{H^2/2k^3}=\frac{\lambda^2}{4} \log(k_2/k_1),
\ee
where the $4\pi$ comes from the angular integration.
It is natural to take $k_2$ to be the scale at which the linearized approximation breaks down, which is roughly $( 5$ Mpc/h$)^{-1}$. At the other end, $k_1$ is bounded by the Hubble scale. Hence the best we can hope for is
\be
\left(\frac{S}{N}\right)^2_{ideal}\approx \frac{3 \lambda^2}{2} ,
\ee
which implies that for the $S/N$ to be larger than one we should have
$\lambda> \sqrt{2/3} $.
Since the dependence of $S/N$  on the size of the survey, $1/k_1$, is only logarithmic, this conclusion is not changed much for realistic surveys with $1/k_1$ of the order of $100$ Mpc/h.

We conclude that  the imprints of a PIP  in a sky survey are detectable if
$| \lambda| \gtrsim 1$.
To actually be able to detect the PIP's imprints we need to know what to look for. This is described in the next sections.

\subsection{Applicability of the approximation}

So far we assumed that there is a clear separation between the effect caused
by the PIP, $\delta\phi_{PIP}$, and the standard power spectrum caused by
the usual quantum fluctuation of the inflaton (\ref{power}). For large $\lambda$
we expect this assumption to break down and  to find mixing between the two.

In other words,  we have neglected the backreaction of $\delta\phi_{PIP}$ which for large $\lambda$  will alter the power spectrum in a significant way.
To estimate when the backreaction of $\delta\phi_{PIP}$ is small we compare it to the standard driving force of the
inflaton, $V^{'}$, 
\be
\frac{1}{a(t)^2}\nabla^2 \delta\phi_{PIP}(r)\ll V',
\ee
where $\delta\phi_{PIP}(r) = \lambda\frac{H}{4\pi}\log(r)$ is the Fourier
transform of  (\ref{PIP}). This condition should be satisfied  at horizon crossing
when the power spectrum is determined. Thus we get
\be
\lambda\ll\frac{4\pi V^{'}}{H^3 }\cong 10^5,
\ee
where we have used the relation $H^2=V/3$ and the COBE normalization ($V^{3/2}/V'=5.169\cdot10^{-4}$). 

The  values of $\lambda$ that appear in the rest of the paper are at most $10^2$ which implies that neglecting the backreaction of $\delta\phi_{PIP}$ is a good approximation.

\sectiono{Large scale structure}

So far we have concluded that a PIP will create a spherically symmetric defect in the inflaton profile. This in turn provides the seed for a spherically symmetric cosmic defect (SSCD) whose properties we study here. We start with the gravitational potential and discuss its behavior in the small and large scale limits (compared to the comoving Hubble scale at matter-radiation equality). We then  discuss the  energy density profile of the SSCD.

We work in the linearized approximation in which the gravitational potential associated with the SSCD is
\be
\Phi_{SSCD} (k,z) = \frac{9}{10}\Phi_0(k)T(k)D_1(z)(1+z),
\ee
where $\Phi_0$ is the primordial gravitational potential induced by the PIP (see Appendix A.2)
\be\label{yuj}
\Phi_0(k)=\sqrt{\frac{2}{9\epsilon}} \delta \phi_{PIP}(k)=
\frac{\lambda}{k^3}
\frac{H}{12\sqrt{\pi\epsilon}},
\ee
and $T(k)$ is the transfer function. The evolution of the potential with the redshift (denoted by $z$) in
$\Lambda$CDM is encoded in $D_1(z)(1+z)$, where $D_1(z)$ is the growth
function defined by
\be
D_1(z)=\frac{5}{2}\Omega_m E(z)\int_{z}^{\infty}\frac{1+z'}{E(z')^3}dz',
~~~~\mbox{with}~~~~ E(z)=\sqrt{\Omega_m(1+z)^3+\Omega_\Lambda}.
\ee

For an actual detection it is useful to know  the real space profile which is obtained by a Fourier transform of $\Phi(k)$.
At distances much larger than the comoving Hubble scale during matter-radiation equality, $\sim110$ Mpc, the transfer function is approximately $1$, which yields\footnote{We assume for simplicity that $n_s=1$ and that $\lambda$ is a constant.}
\be\label{e}
\Phi_{SSCD}(r,z) = - \lambda\frac{\sqrt{3}}{20\pi}\frac{V^{3/2}}{V'} D_1(z)(1+z)\log\left(r\right).
\ee We denote by $r$ the distance from the center of the SSCD and measure it in units of Mpc/h.
Using the COBE normalization ($V^{3/2}/V'=5.169\cdot10^{-4}$) and setting $z=0$ we find
\be\label{largedis}
\Phi_{SSCD}(r, z=0)= |\lambda|C\log\left(r\right),
\ee
where
\be
C =\pm 1.09 \cdot 10^{-5}.
\ee
The plus (minus) sign is obtained when $\lambda V^{'} <(>)~0$ for which (\ref{largedis}) describes an over (under) dense spherically symmetric region.  If $m$ does not depend on the inflaton then $\lambda= -\frac12 V^{'}m/V$ and $\lambda V^{'} < 0$. Therefore, a PIP with a constant mass provides the seed for an over dense region. For a PIP to induce an under dense region (or a void), its $\lambda$ should be dominated by $m^{'}$ with $m^{'} V^{'}>0$. We shall use this large distances approximation throughout the paper, when relevant.

\begin{figure}\label{fig:PV}
\begin{picture}(220,150)(0,0)
\vspace{0mm} \hspace{27mm} \mbox{\epsfxsize=100mm
\epsfbox{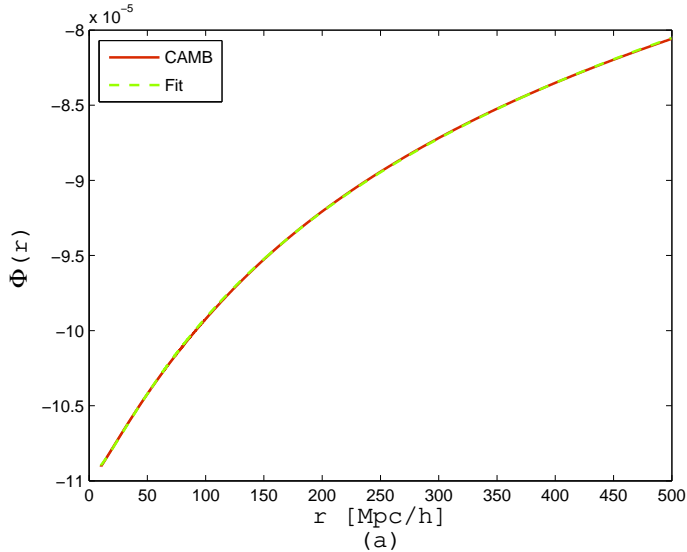}}
\end{picture}
\caption{The profile of the gravitational potential $\Phi_{SSCD}(r)$  for $\lambda = 1$. The {\it solid red} line stands for the potential $\Phi_{SSCD}(r)$, calculated numerically (the transfer function was calculated via CAMB). We fit the potential numerically in the region $10$ Mpc/h $<r<500$ Mpc/h and the result $\Phi_{fit}(r)$ is plotted with a {\it dashed green} line.}
\end{figure}

At distances shorter than $110$ Mpc, the transfer function is  more complicated and to obtain an accurate profile of the potential one has to rely on numerical methods.
To this end we used a standalone version of CAMB \cite{CAMB} with cosmological parameters taken from the joint results of WMAP+BAO+SNI  \cite{WMAPcosmo}:
 $ H_0 = 70.1$ km/sec/Mpc, $\Omega_b = 0.0462$, $\Omega_c = 0.233$ and $\Omega_\Lambda = 0.721$.
We examined the resulting potential in the range $10$ Mpc/h $<r<500$ Mpc/h and found a good fit to the numerical curve. The fitting function is
\be\label{Gfit}
\Phi_{fit}(r)=\lambda\cdot10^{-5}   \left(a\frac{r}{b+r}\log(1+\frac{r}{c})+d\right),
\ee
where the numerical values of the parameters are
\ben
&& a =1.1394\pm0.0023,~~~~~~b = 20.27 \pm 0.23 ~\mbox{Mpc/h},\nonumber\\
&&c = 75.44 \pm
0.36~\mbox{Mpc/h},~~~~~~d = -0.8392 \pm0.0038 .
\een
Figure 1 illustrates how well (\ref{Gfit}) approximates the CAMB result in this range.

The energy density associated with the SSCD is determined by the gravitational potential in the standard way
\be\label{d}
\delta(r)=\frac{1}{4\pi G \rho_m} \nabla^2\Phi(r,z=0).
\ee
At distances much larger than $1/k_{eq} \sim 110$ Mpc, eq. (\ref{largedis}) is applicable and gives
\be\label{pa}
\delta(r)\cong \lambda\frac{233 }{r^2}.
\ee
Therefore, at large distances the SSCD has a profile identical to that of an isothermal sphere, and in particular  $M(r)$  grows linearly with the radius $M(r)\sim r$.

As expected, the SSCD has a stronger effect at large scales than the standard large scale structure, which  decays faster with the distance. An example is the well-known NFW profile \cite{NFW} where $\delta(r)\propto r^{-3}$ at large distances. This, together with the fact that it is spherically symmetric (as opposed to a typical cosmic web that includes filaments, walls and voids) suggests that if indeed  SSCDs are present in the visible universe (and if $|\lambda |$ is large enough), one should be able to tell them apart from the standard large scale structure.

At shorter distances we can combine  (\ref{Gfit}) and (\ref{d}) to find the density profile.
Expanding around $r=0$, we get
\be
\rho(r)=\lambda\left( 0.96 -0.11 r +0.009 r^2 +...\right) .
\ee
We see that for $|\lambda |$ of the order of $1$ the core of the SSCD involves non linear effects.

\sectiono{CMB imprints}

In this section we study the imprints of a SSCD in the CMB.
Our main goal is to calculate the CMB signal to noise ratio ($S/N$) associated with a single SSCD as a function of the only two free parameters: its magnitude, which depends on $\lambda$,  and distance from the observer which we denote by  $r_0$. The dependence of the signal on $\lambda$ is linear. However, the dependence on $r_0$ is quite interesting and it involves some interesting physics.

In the first subsection we focus on  the Sachs-Wolfe (SW) effect \cite{SW}. Calculating the SW signal associated with the SSCD can be done analytically and it illustrates some key features.  For example, we show that, as expected from such a SSCD,  the main contribution to the $S/N$ comes from the low-$l$ modes. This means that the late integrated SW (ISW) effect cannot be neglected.
In the second subsection we add the ISW effect associated with the SSCD  and calculate the combined SW+ISW signal numerically. We find that the combined  signal has an interesting dependence on $r_0$.  Its main feature is a dip in the signal at around $r_0 \sim 4700$ Mpc/h.

\subsection{SW signal}

To calculate the anisotropy we  work in the following coordinate system (see Figure 2). We take the z-axis (not to be confused with the notation for redshift $z$) to point towards the center of the SSCD. The distance between the SSCD and the observer is denoted by $r_0$  and is measured in units of Mpc/h. The last scattering surface is a sphere of radius $r_{lss}$ from the observer . In principle $r_0$ can be larger than $r_{lss}=9750$ Mpc/h. Practically, however, this case is not very interesting since it yields a small $S/N$.

In this coordinate system the gravitational potential associated with the SSCD is
\be\label{qa} \Phi_{SSCD}(\vec r,z) = \lambda\tilde{C}D_1(z)(1+z)\log(|\vec
r-\vec r_0|), \ee
where $\tilde{C}=\frac{C}{D_1(z=0)}=1.425\cdot 10^{-5}$ and  $|\vec r-\vec
r_0| =\sqrt{r^2+r_0^2-2rr_0\cos\theta}$.
\begin{figure}\label{fig:LSS}
\begin{picture}(200,150)(0,0)
\vspace{0mm}\hspace{45mm}\mbox{\epsfxsize=90mm
\epsfbox{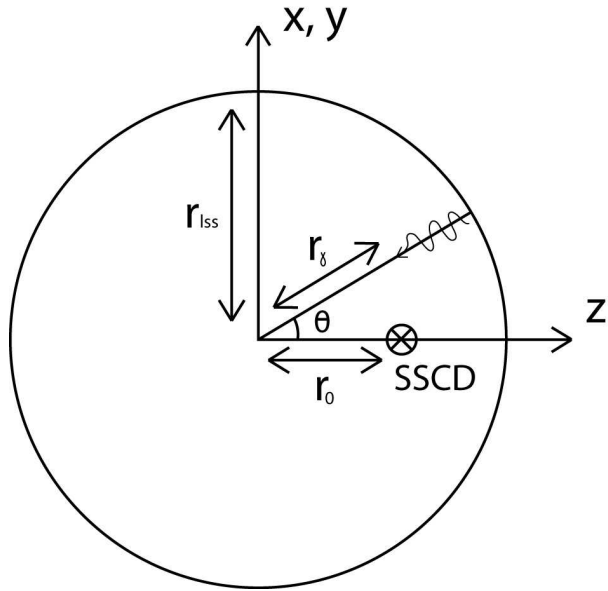}}
\end{picture}
\caption{The basic setup. The $z$-axis  points towards the center of the SSCD. The distance between the SSCD and the observer is denoted by $r_0$. Due to the rotational symmetry of the setup we need to specify only one angle defined with respect to the z-axis, which we denote by $\theta$. The distance between the observer and the photon along its path we denote by $r_{\gamma}$ and the last scattering surface is located at a distance $r_{lss}$ ({\it lss} stands for last scattering surface). }
\end{figure}
The SW temperature anisotropy associated with the SSCD reads
\be\label{SW2}
\frac{\delta T^{SW}}{T}=  \frac{\Phi_{SSCD}(r_{lss},z_{lss})}{3}=\frac{1}{6}\lambda\tilde{C}
\log(r_{lss}^2+r_0^2-2r_{lss}r_0\cos\theta),
\ee
where we assume that the decoupling occurs during the matter dominated epoch, which means that $D_1(z_{lss})(1+z_{lss})=1$. In (\ref{SW2}) we ignore the monopole term ($\Phi_{SSCD}(0)/3$) which does not yield any interesting effects.

To calculate the $S/N$  we decompose  the temperature anisotropy into spherical harmonics
\be\label{dec}
\frac{\delta T}{T}(\theta,\phi)=\sum_{l=1}^{\infty}\sum_{m=-l}^{m=l} a_{lm} Y_{lm}(\theta,\phi).
\ee
The fact that the potential profile (\ref{qa}) has azimuthal symmetry  means that we get a non-vanishing signal only for $m=0$. In other words, all the information about the SSCD is in the one-point function of the $m=0$ modes
\be\label{sig}
S_l\equiv \langle a_{l,m=0} \rangle.
\ee

Combining (\ref{SW2}), (\ref{dec}) and (\ref{sig}) we find  the  SW   signal to be
\be\label{lk}
S_l^{SW}=\lambda\tilde{C}\frac{\sqrt{\pi(2l+1)}}{6} \int^1_{-1} dx
P_l(x)\log\left( 1-2yx+y^2 \right), \ee
where  $y \equiv r_0/r_{lss}$ and $x\equiv\cos\theta$.
The full SW signal to noise ratio is
\be\label{rd}
\left( \frac{S^{SW}}{N}\right)^2=\sum_{l=2}(S_l^{SW})^2/C_l,
\ee
where $C_l$ is determined by the  two-point function associated with the standard effects of $\Lambda$CDM
\be
\langle a_{lm} a^{*}_{l^{'} m^{'}}\rangle=\delta_{l l^{'}} \delta_{m m^{'}} C_l.
\ee
As usual, the $l=1$ mode is ignored  since it is mixed with the Doppler effect. We will return to this mode in section 5.

For each $l$,  (\ref{lk}) can be computed analytically. For small $y$ we can expand the result to find that the leading term scales like \be \label{yf} S^{SW}_l\propto -\lambda y^l/\sqrt{l},\ee which yields a $S/N$ that decays with $l$
\be
 (S_l^{SW})^2/C_l\sim  \lambda^2 y^{2l}l.
\ee
Thus, as expected, the main contribution to (\ref{rd}) for small $y$ ($r_0 \ll  r_{lss}$) comes from the low-$l$s. With a bit of  work it can be shown that the leading contribution comes from the low-$l$ modes for any  $y$.

\subsection{SW+ISW signal}

The fact that the main contribution to the $S/N$ comes from the low-$l$ modes means that it is not a good approximation to neglect  the contribution of the ISW effect  to the signal.  Namely, we have to consider the combined $S/N$
\be
\left( \frac{S^{SW+ISW}}{N}\right)^2
=\sum_{l=2} \left( S_{l}^{SW}+S_{l}^{ISW}\right) ^2 /C_l.
\ee
Since the  SW and ISW imprints of a large scale structure on the CMB have an opposite signature\footnote{An over dense region creates a cold spot when the ISW effect dominates and a hot spot when the SW effect dominates.},  one might suspect that the competition between the two could lead to some interesting effects. As we show below, this is indeed the case.

Decomposing the ISW effect associated with the SSCD
\be
\frac{\delta T^{ISW}(\theta)}{T}=2\int^{\tau_0}_{lss}\frac{\p \Phi_{SSCD}(\theta,\tau)}{\p\tau}d\tau
\ee
into spherical harmonics, we find (using (\ref{dec}) and (\ref{sig}))
\ben \label{ISW}
&& S_{l}^{ISW}=\\
&&- \lambda\tilde{C}\sqrt{\pi (2l+1)}\int^{z_{lss}}_0 dzD_1(z)
\left(1-f(z)\right)\int^1_{-1}dxP_l(x)\ln(1+y_{\gamma} (z)^2-2y_{\gamma}(z)x),\nonumber
\een
where
\be
f(z)= -\frac{1+z}{D_1(z)}\frac{\p D_1(z)}{\p z},
\ee
$r_{\gamma} (z)$ is  the position of the photon along its trajectory from the surface of last scattering to the observer and
$y_{\gamma}(z)= r_{\gamma} (z)/r_0$ (see Figure 2).

For  quantitative results we have to rely on numerical calculations. There is, however, an important qualitative feature of this integral that can be deduced analytically.
Along the position of the photon there is always a region in which $r_{\gamma}\sim r_0 $. In this region the integral over $x$ in (\ref{ISW}) is of order $1$. Therefore, for small $y$
\be\label{yg}
S_{l}^{ISW}\propto \lambda y,
\ee where again $y=r_0/r_{lss}$.
Comparing (\ref{yg}) with (\ref{yf}) we see that for small $y$ the leading contribution comes from the ISW effect\footnote{$y$ cannot be too small since all along we are using the long range approximation result (\ref{qa}) which is valid for $y> 110/r_{lss}\sim 10^{-2}$.}.

As we increase $y$, the ISW effect decays. This happens because  in this case a CMB photon passes through the SSCD region when the universe is still dominated by cold matter. Since in a matter dominated universe the ISW effect vanishes, the ISW contribution arises from a small tail of the potential which the photon experiences at later times. Therefore at large $y$ the contribution of the ISW decreases with $r_0$.  Simultaneously, the anisotropy due to the SW effect grows as the SSCD approaches the last scattering surface. At a certain point the ISW  becomes negligible compared to the SW. Since
$S^{SW}$  and $S^{ISW}$ have opposite signs, for each $l$ there exists a point, denoted by $r_l$, where the two effects cancel each other
\be
S_{l}^{SW}(r_0=r_l)+S_{l}^{ISW}(r_0=r_l)=0.
\ee
Similar reasoning implies that $r_l$ grows with $l$.

Numerical calculations of $S_l^{SW}(r_0)$ and $S_l^{ISW}(r_0)$ show that indeed both decay with $l$ and that for any practical purpose it is enough to account for the first $50$ multipoles. In addition, the numerical calculation shows that  indeed $r_l$ grows with $l$ and that for low-$l$, $r_l$ depends fairly mildly on $l$. For example,  in Figure 3 (a) we plot the lowest multipoles of the SW+ISW signal versus $r_0$ and show that
\be
r_2 = 4400~ \mbox{Mpc/h}, ~~~ r_3 = 4700 ~\mbox{Mpc/h}~~~\textrm{ and} ~~~r_4 = 4980 ~\mbox{Mpc/h}.
\ee
Since the main contribution to the signal comes from the low-$l$ modes, the SW-ISW cancellation leads to a dip in the $S/N$ around the low-$l$ cancellation region, i.e. around $r_0=4700$ Mpc/h (we will discuss this feature later).

\begin{figure}\label{fig:SWISW}
\begin{picture}(220,150)(0,0)
\vspace{0mm} \hspace{0mm} \mbox{\epsfxsize=85mm
\epsfbox{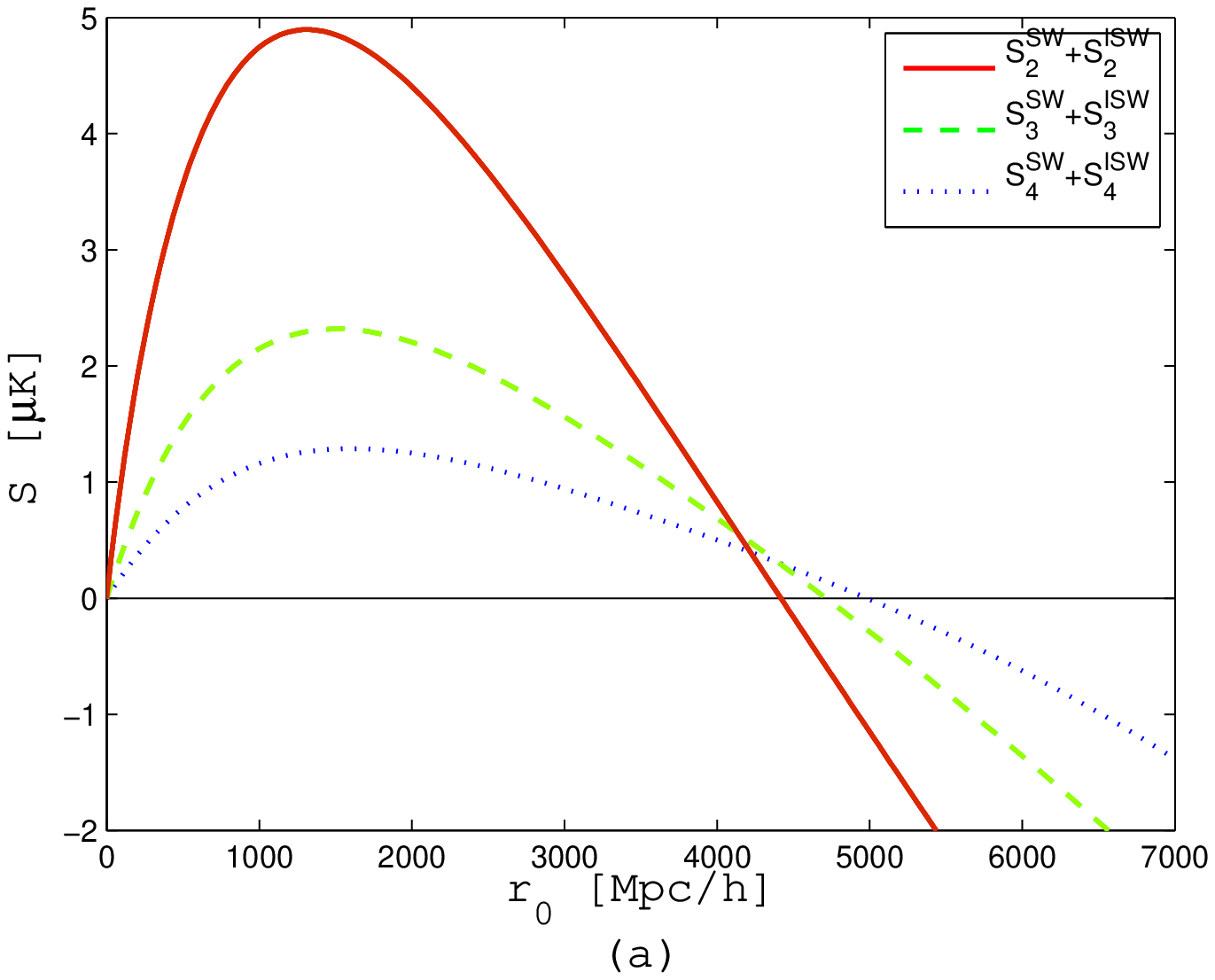}}
\end{picture}
\begin{picture}(220,150)(0,0)
\vspace{0mm} \hspace{0mm} \mbox{\epsfxsize=85mm
\epsfbox{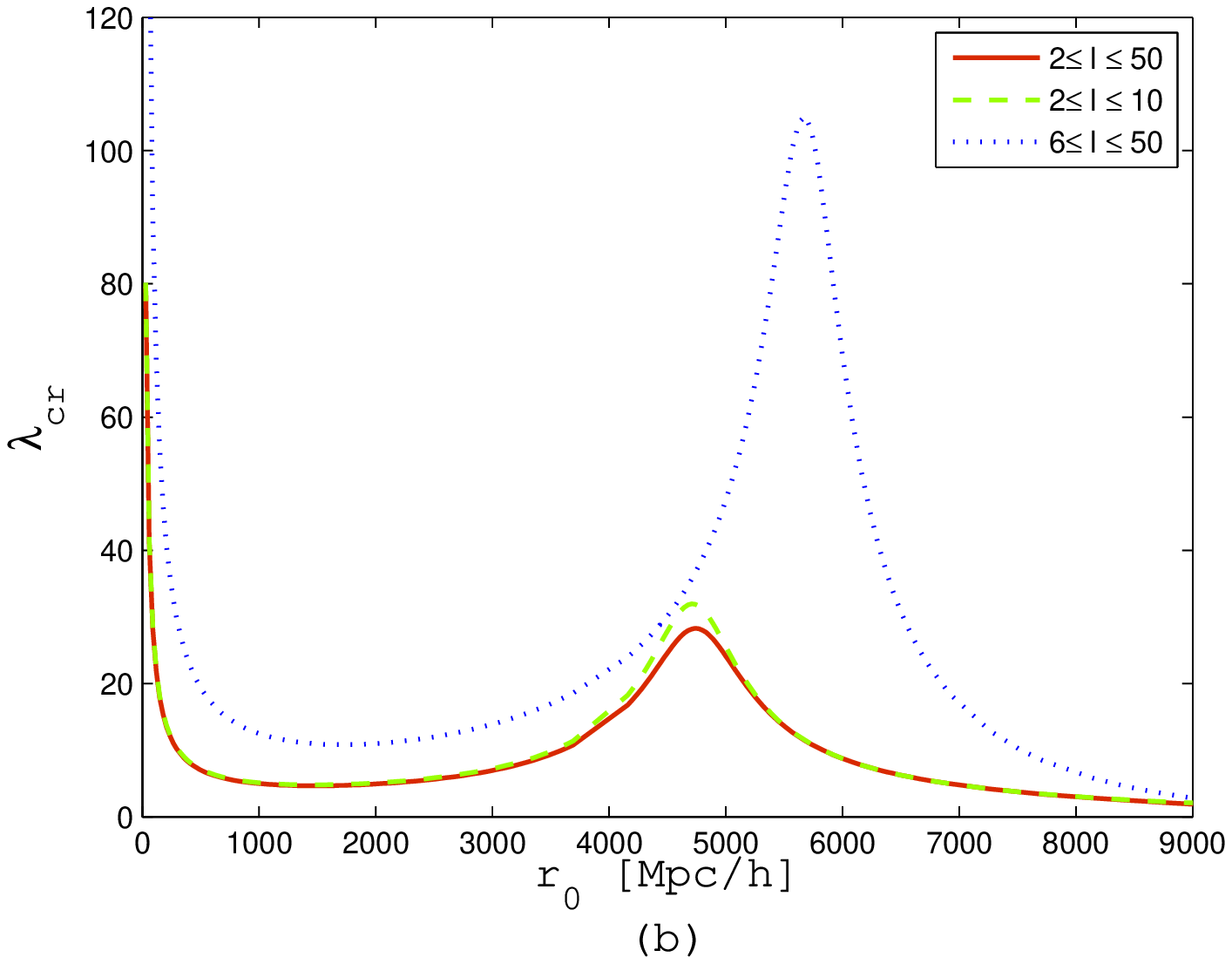}}
\end{picture}
\caption{In (a) we plot the signal $(S_l^{SW}+S_l^{ISW})$ in $\mu K$ versus $r_0$ for $\lambda = 1$. We demonstrate the cancellation of the SW and ISW effects for different multipoles: quadrupole ({\it solid red}), octupole ({\it dashed green}) and $l=4$ ({\it dotted blue}). For different $l$ the cancellation occurs at different $r_l$:  $r_2 = 4400$  Mpc/h, $r_3 = 4700$ Mpc/h and $r_4 = 4980$ Mpc/h. Note that $r_l$  grows (slowly) with the multipole number. In (b) we plot the critical parameter $\lambda_{cr}(r_0)$ versus $r_0$ for different multipole ranges:  $2\leq l \leq 50$ ({\it solid red}), $2\leq l \leq 10$ ({\it dashed green}) and $6\leq l \leq 50$ ({\it dotted blue}). This illustrates that the main effect comes from lowest multipoles. In order to account for higher multipoles we should have large values of $\lambda_{cr}(r_0)$.}
\end{figure}

Our next goal is to pose constrains on the parameters $\lambda$ and $r_0$.
In particular, we wish to know for what range of the parameter $\lambda$ at every $r_0$ the SSCD is observable in a CMB experiment.
 As we discussed earlier, the signal scales with $\lambda$.
Therefore, for every $r_0$ there is a critical value of $\lambda$, denoted by $\lambda_{cr} (r_0)$, such that for
$\lambda> \lambda_{cr} (r_0)$ the signal is larger than the noise. Namely, for $\lambda$ larger than its critical value, the imprints
of the SSCD in the CMB are detectable. In Figure 3 (b) we plot the critical value at each location of the SSCD. We also study the contribution of the signal in different multipoles.

For small $r_0$ the signal is dominated by  the ISW effect, which is proportional to $r_0$ as was shown earlier. Thus the critical parameter behaves as
\be
\lambda_{cr}(r_0)\propto r_{lss}/r_0.
\ee
As we increase $r_0$ the ISW starts to deviate from the linear dependence on $r_0$ and the SW  effect begins  to suppress the ISW effect. As a result there is a minimum to $\lambda_{cr}(r_0)$ at around $r_0 = 1500$ Mpc/h where its value is   $\lambda_{cr}(r_0=1500$ Mpc/h$)=4.7$.

When we increase $r_0$ further the SW-ISW cancellation becomes more effective and there is  a fairly clear peak
at around $r_0=4700$ Mpc/h when we account for all the information (i.e. for  $2\leq l\leq 50$ multipoles).
Note that the critical parameter  is almost identical when we keep only the contribution of $2\leq l \leq  10 $. This illustrates that the main signal comes from the low-$l$ modes. The very low-$l$ modes, $l\leq 5$, are somewhat anomalous (for a recent discussion see  \cite{Copi:2008hw}), and so it is natural to calculate the $S/N$ excluding these modes. As Figure 3 shows, this leads to two main effects. First, since the signal is considerably weaker,  $\lambda_{cr}(r_0)$ is significantly larger. Second, the peak is shifted to around $r_0 = 5700$ Mpc/h, which is in the cancellation region of $l=6$, $l=7$ and $l=8$ ($r_6 = 5400$ Mpc/h, $r_7 = 5600$ Mpc/h and $r_8=5770$ Mpc/h respectively).

\begin{figure}\label{fig:Hs}
\begin{picture}(220,150)(0,0)
\vspace{0mm} \hspace{0mm} \mbox{\epsfxsize=85mm
\epsfbox{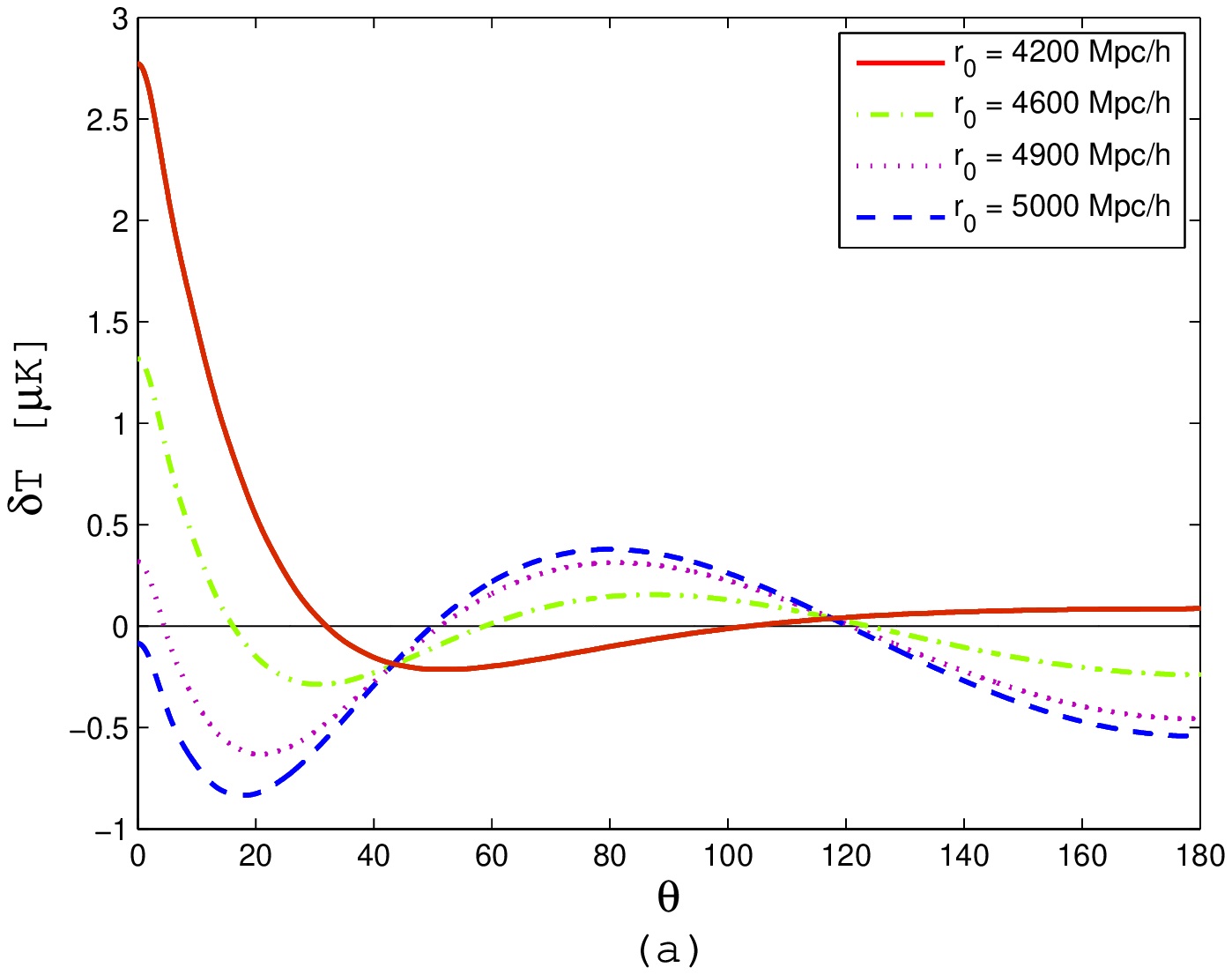}}
\end{picture}
\begin{picture}(220,150)(0,0)
\vspace{0mm} \hspace{0mm} \mbox{\epsfxsize=85mm
\epsfbox{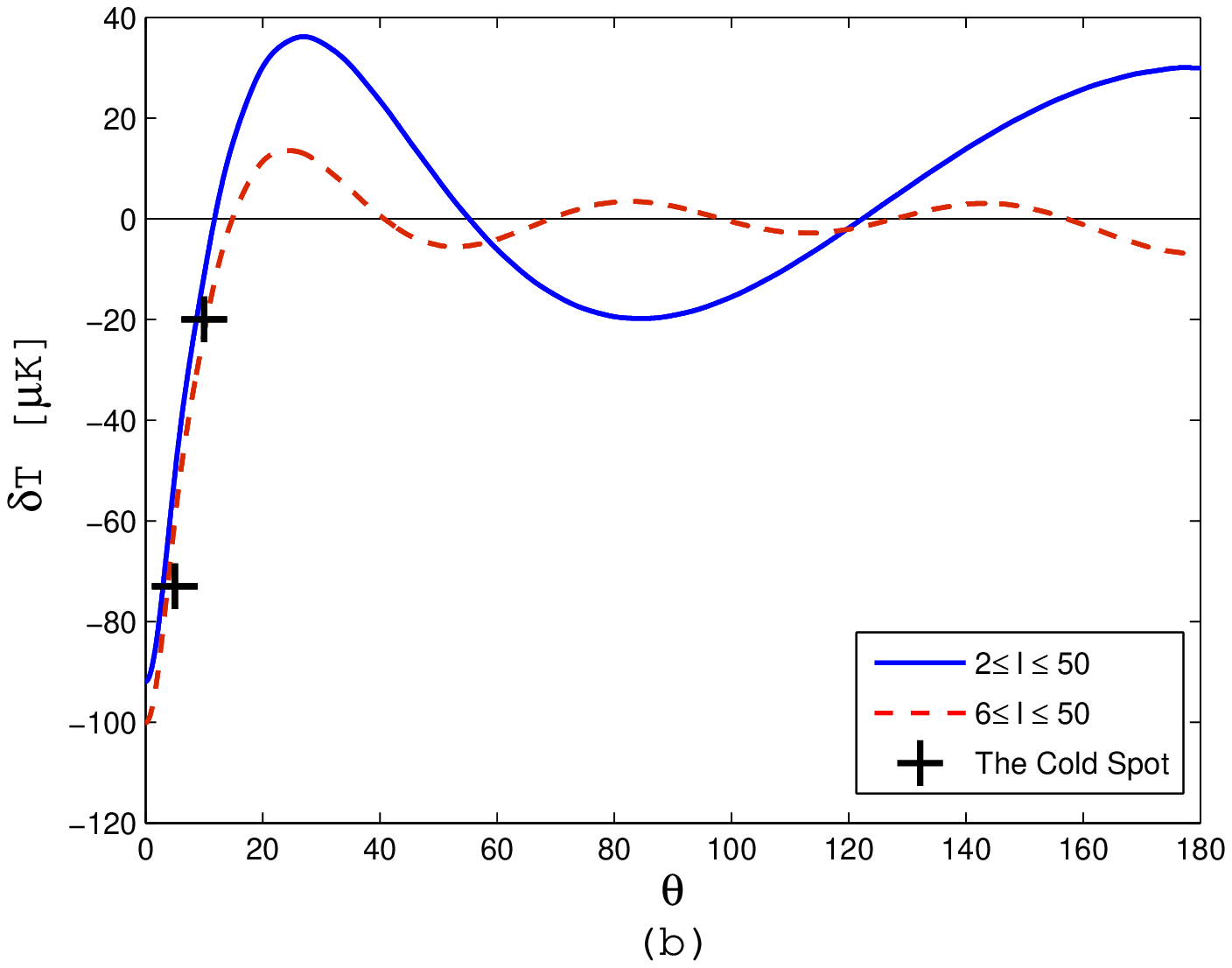}}
\end{picture}
\caption{   (a) A  hot spot created by a SSCD with $\lambda = 1$ in the range $r_0\sim 4200$ Mpc/h $ -~5000$ Mpc/h.  The radius of the spot decreases while $r_0$ increases. In (b) we demonstrate an attempt to relate the SSCD to the WMAP Cold Spot. We plot two published temperature measurements of the Cold Spot (in {\it blue pluses}) and fit it with a profile created by a SSCD of $\lambda\sim -95$ and located at $r_0\sim4700$ Mpc/h ({\it solid blue} for $2\leq l\leq 50$ and $6\leq l \leq 50$ in {\it dashed red}).}
 \end{figure}

The SW-ISW cancellation leads to another interesting effect that can help in detecting the SSCD imprints.
In the range $4200$ Mpc/h $<r_0<5000$ Mpc/h, ($2.55< z <3.75$) the temperature profile  associated with the SSCD is highly sensitive to its location, $r_0$.
A SSCD located in this range manifests itself as a {\it localized} hot or cold spot in the CMB temperature map, depending on the sign of $\lambda$.
A rough explanation for this is the following:
as we increase $r_0$, more and more low-$l$ multipoles approach the cancellation region $r_l$ and thus almost do not contribute to the signal. Therefore, in this range the signal is dominated by some specific multipole $l$ and this multipole number grows with distance. As a consequence, the  characteristic scale of the temperature profile becomes smaller.
  In other words, the larger  $r_0$ is, the smaller the spot is.

Therefore, the radius of such a spot can be used to determine $r_0$. This is illustrated in Figure 4 (a) where we have plotted the  temperature profile
\be
\frac{\delta T}{T}(\theta)=\sum_{l=2}S^{SW+ISW}_{l}Y_{l0}(\theta),
\ee
associated with a SSCD with $\lambda=1$ for different values of $r_0$ in this range.
We see that throughout this region the profile is dominated by fairly small angles and that it is quite sensitive to $r_0$. For $r_0=4200$ Mpc/h, the radius of the hot spot is about $32^{\circ}$. For $r_0=4600$ Mpc/h, the radius of the hot spot is considerably smaller, about $16^{\circ}$, and the amplitude is weaker. For $r_0=4900$ Mpc/h the radius of the spot is tiny, $4.5^{\circ}$, and the amplitude is so small that it is overshadowed by a cold ring that peaks at about  $20^{\circ}$. By the time we reach $r_0=5000$ Mpc/h, the ring swallows the tiny hole and we have a  fairly large cold spot of radius $45^{\circ}$.

The WMAP cold spot  \cite{Vielva, Cruz04, Cruz06a, Cruz06b} might be explained in this fashion. The WMAP cold spot is a nearly spherically symmetric region with an approximate temperature  $\delta T = -73~\mu K$ at  $\sim 5^{\circ}$ \cite{Cruz04} and an average temperature, $\delta T\sim -20~\mu K$ at angular radius of $\sim 10^{\circ}$, \cite{Cruz06b}. This fits well with the imprints of a SSCD with $\lambda\sim -95$ located at $r_0\sim 4700$ Mpc/h (see Figure 4 (b)). In our case the Cold Spot is surrounded by a  hot ring with a peak at about  $\sim 30^\circ$. This, as well as improved measurements (perhaps from ACT \cite{ACT} and SPT \cite{SPT}), might be used to distinguish the SSCD from other possible explanations, such as a localized void \cite{Rudnick, Inoue:2006fn, Inoue:2006rd} and a cosmic texture
\cite{Cruz:2007pe}. Unfortunately, the fact that the magnitude and shape of the hot ring, unlike the cold-spot, is quite sensitive to the very low-$l$ modes ($2\leq l\leq 5$) makes it less trustable.

\begin{figure}\label{fig:LSS}
\begin{picture}(220,150)(0,0)
\vspace{0mm}\hspace{27mm}\mbox{\epsfxsize=100mm
\epsfbox{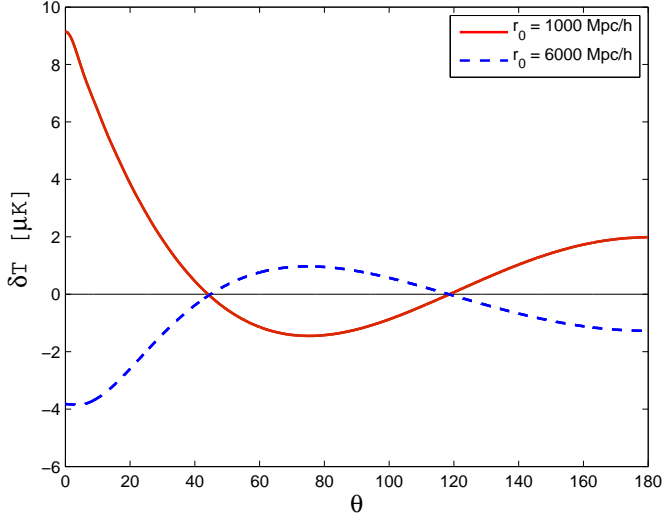}}
\end{picture}
\caption{
 We show the pattern of $\delta T(\theta)$ created by a nearby ({\it solid red}) and remote ({\it dashed blue}) SSCD, both of $\lambda = 1$. The former creates a diffused hot spot (the ISW effect is dominant) and the latter produces a diffused cold spot (the SW effect is dominant).}
\end{figure}

Outside the SW-ISW cancellation region  $4200$ Mpc/h $<r_0<5000$ Mpc/h, the main signal is due to the low-$l$ modes. Therefore, the SSCD signature at these locations is a mild modulation of the temperature anisotropy and has no prominent pattern (see Figure 5). Such a signature is not easily detectable in the real-space anisotropy map of the CMB but can be seen in its decomposition in spherical harmonics.

\sectiono{Bulk flow imprints}

Another interesting large scale effect caused by the SSCD is the peculiar velocity field it induces around it. The peculiar velocity is determined by the gravitational potential
\be \label{PecV}
\vec{v}(z,\vec r)=-\frac{2}{3}\frac{H(z)}{(1+z)^2H_0^2\Omega_m}f(z)\vec\nabla\Phi(z,\vec r),
\ee
and it points towards (away from) an over (under) dense region.
In this section we calculate the bulk flow associated with a SSCD and determine when it  dominates the rms bulk flow predicted by the  $\Lambda$CDM model.
The main motivation for this comes from analyses such as \cite{Watkins,Lavaux:2008th} that reported an anomalously large bulk flow.

The observed dipole has two contributions
\be
D^{observed}=D^{Doppler}+D^{Gravity},
\ee
both determined by the gravitational potential and can be straightforwardly calculated in our setup.
The first contribution is due to the velocity of the local bulk relative to the last scattering surface. Hence we get
\be
D^{Doppler}=\sqrt{3\pi}\int^1_{-1}dxx\frac{\Delta T^{Doppler}(x)}{T},
\ee
with
\be
\frac{\Delta
T^{Doppler} (\hat n)}{T} = \hat n\cdot(\vec v_{bulk}-\vec
v_{lss}(\hat n)).
\ee
In our case we  use the gravitational potential generated by the SSCD (\ref{qa}) and (\ref{PecV})  to obtain
\ben
&& \hat n \cdot\vec v_{bulk} =
\lambda\frac{\tilde C}{H_0r_{lss}}\left(\frac{5}{3}-D_1(0)\right)\frac{x}{y},\nonumber \\
&& \hat n \cdot \vec
v_{lss}  =
-\lambda\frac{2\tilde C}{3\sqrt{z_{lss}}r_{lss}}\frac{(1-y x)}{1+y^2-2y x}.
\een
Note that our local motion yields a pure dipole
term whereas the motion of the last scattering surface induces
higher order multipoles as well. We did not account for them in the previous
section since their contribution is suppressed by a factor of $1/\sqrt{z_{lss}}$
compared  to the SW and ISW effects.
Similarly the leading contribution to $D^{Doppler}$ is due to the local bulk motion and not the motion of the last scattering surface
\be
D^{Doppler} \cong   1.84\frac{\lambda\tilde C}{H_0r_{0}}.
\ee

The second contribution to the bulk flow comes from the gravitational potential the CMB experiences. Namely it is the $l=1$ modes of the CMB anisotropy, given at leading order by
\be
D^{Gravity}=S_1^{SW}+S_1^{ISW}.
\ee
In our case $S_1^{SW}$ and $S_1^{ISW}$ are  determined by (\ref{lk}) and (\ref{ISW}) respectively.
As follows from the discussion in the previous section, both $S_1^{SW}$ and $S_1^{ISW}$ scale like $y$ in the limit of small $y$ and so $D^{Gravity}$ is negligible  compared to $D^{Doppler}$ in this limit.
Therefore, at small distances (compared to the Hubble scale) we have
\be\label{dd}
v_{observed} \cong
0.0385\frac{\lambda}{r_0}.
\ee
Recall that we are working in units where $c=1$=Mpc/h.

Next we consider the rms bulk flow associated with a region of radius $R$ predicted by the  $\Lambda$CDM model (for a review see \cite{Strauss}). For large $R$ we are in the linearized region and the average bulk flow reads
\be 
v_{rms} 
\cong\frac{0.0183}{R}.
\ee
For example, for $R=50$ Mpc/h (which is relevant for \cite{Watkins}) we get $v_{rms} \cong 110$ km/sec.

Comparing this with (\ref{dd}) we find that the peculiar velocity signal of a SSCD is larger than the $\Lambda$CDM noise if
\be
\lambda> 0.475 \frac{r_0}{R}.
\ee
Note that our estimate is valid for $r_0\gtrsim2R$ and so the signal is larger than the noise only for $\lambda\gtrsim1$.

\sectiono{Peculiar velocity and CMB}

In this section we combine the peculiar velocity imprint and  the CMB imprints of a single SSCD and argue that quite generically the two are correlated. We also point out a possible experimental hint for such a correlation.

The first step is to find the relationship between $\lambda$ and $r_0$ assuming that a single SSCD is responsible for the anomalously large bulk flow. We use the value reported in \cite{Watkins}, $v_{bulk}\sim 407$ km/sec, and for small $y$ we use (\ref{dd}) to find
\be\label{po}
\lambda^{PV}(r_0) \cong \pm0.0352r_0,
\ee
where as before $r_0$ is measured in units of Mpc/h.
At larger distances we have to take into account the contribution of $D^{Gravity}$ as well. As in the previous section, we use an analytic calculation of the SW contribution and add the ISW contribution numerically.  As expected, $D^{Gravity}$ has an opposite sign compared to $D^{Doppler}$ and so it lowers the magnitude of $D^{Observed}~$\footnote{This is expected since we know that when the cosmic defect is well outside the horizon ($r\gg r_{lss}$) its main effect is to generate a superhorizon mode. The contribution of a superhorizon mode  to $D^{Observed}$ is known to vanish \cite{Turner, Erickcek}. In appendix B we explicitly check that this is indeed the case.}.
In Figure 6 (a) we plot $\lambda^{PV}(r_0)$ (in dashed-dotted purple). The linear behavior at small distances  is in accord with  (\ref{po}), while at larger distances $\lambda^{PV}(r_0)$ grows faster, as a result of the partial cancellation between $D^{Doppler}$ and $D^{Gravity}$.

Next, we wish to study the CMB imprints associated with a SSCD with
$\lambda=\lambda^{PV}(r_0)$.
In particular, we would like to know for which values of $r_0$ the CMB imprints associated with the SSCD are detectable, assuming it is responsible for the bulk flow.
The answer is presented in Figure 6 in two ways. In (a) we plot $\lambda_{cr}(r_0)$ in addition to $\lambda^{PV}(r_0)$. For $r_0$ in which $\lambda_{cr}(r_0)> \lambda^{PV}(r_0)$, the SSCD can induce the large peculiar velocity without leaving a significant imprint in the CMB. For $r_0$ such that $\lambda_{cr}(r_0)< \lambda^{PV}(r_0)$ its CMB imprints are detectable.

Figure 6 (b) presents this in a more quantitative fashion. The CMB $S/N$ is plotted for a SSCD with  $\lambda=\lambda^{PV}(r_0)$ in two cases. In one we include the contribution of all modes $(l\geq 2)$ to the $S/N$. In the second we consider the contributions of only the more reliable modes, those with $6\leq l \leq 18$.  The reason for the lower bound was discussed in the previous section. The upper bound is due to the fact that the bulk flow direction is close to the galactic plane (e.g. $(l,b)=(287^\circ \pm 9^\circ,8^\circ\pm 6^\circ)$ in galactic coordinates according to \cite{Watkins}). Ignoring modes  with $l>18$ (which corresponds to angles under $10^{\circ}$) significantly reduces the sensitivity to the galactic noise.
For this reason we refer to these modes as safe modes.

\begin{figure}\label{fig:PV}
\begin{picture}(220,150)(0,0)
\vspace{0mm} \hspace{0mm} \mbox{\epsfxsize=85mm
\epsfbox{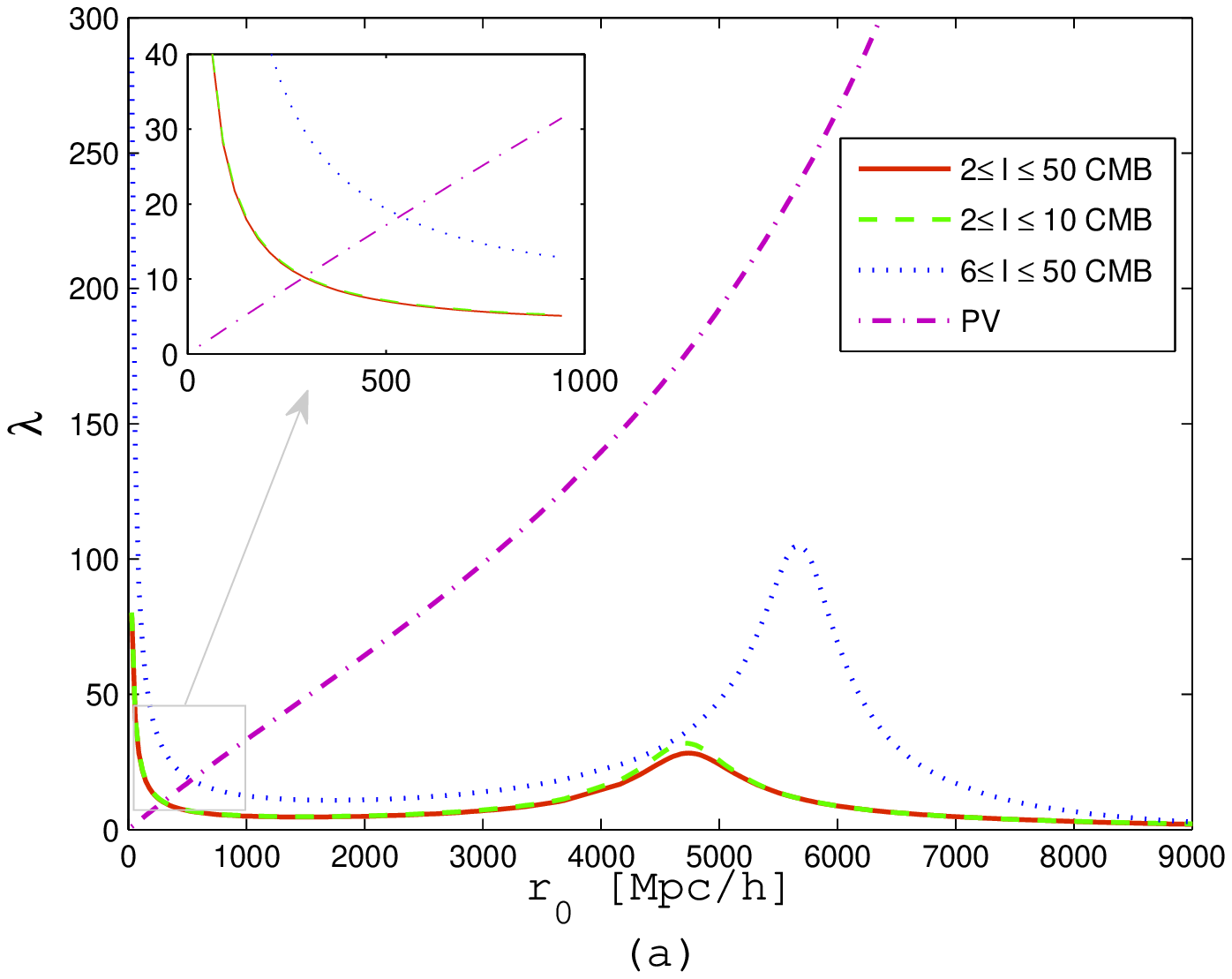}}
\end{picture}
\begin{picture}(220,150)(0,0)
\vspace{0mm} \hspace{0mm} \mbox{\epsfxsize=85mm
\epsfbox{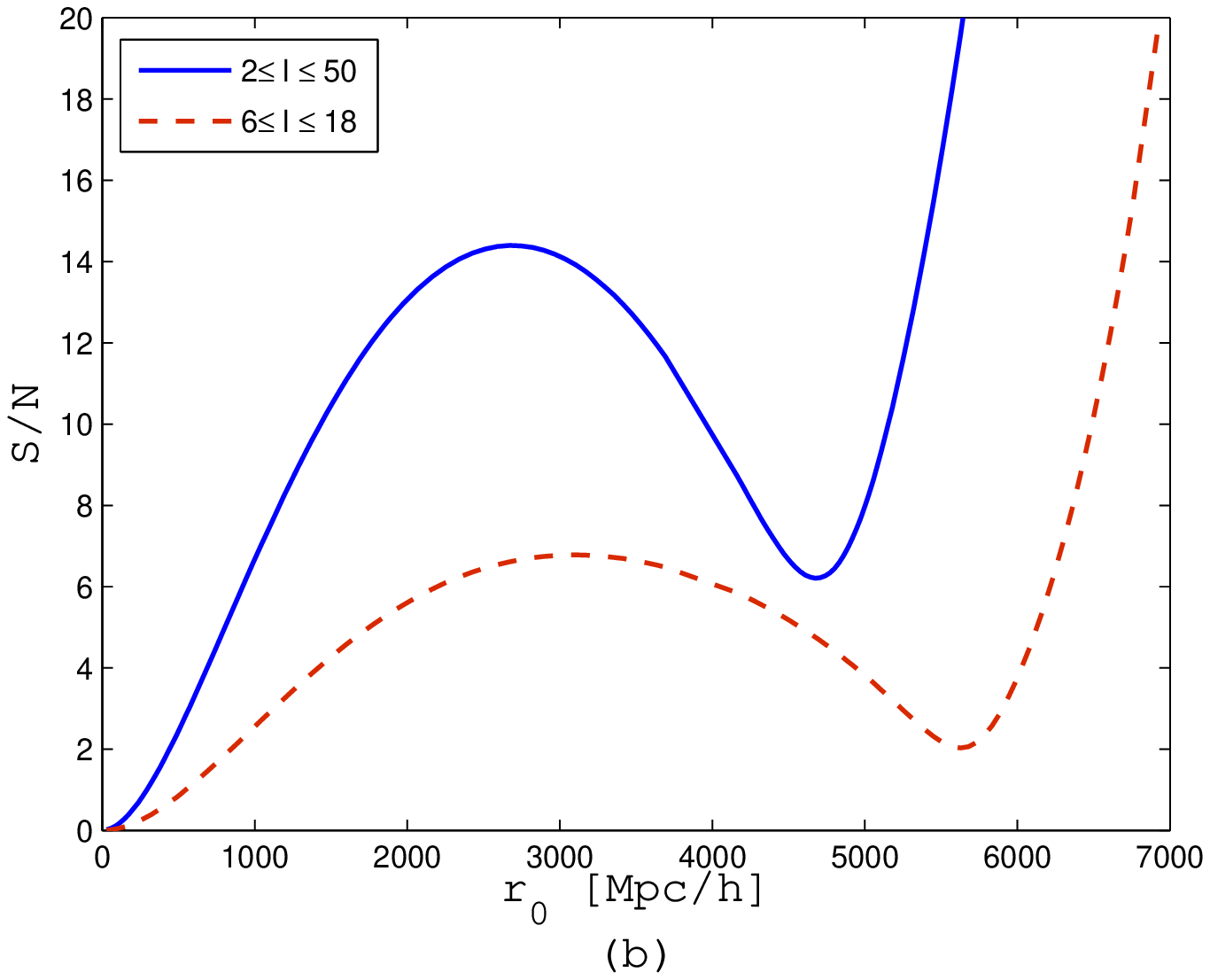}}
\end{picture}
\caption{(a): The critical parameter $\lambda_{cr}(r_0)$ that gives $S/N = 1$ for different multipole ranges ($2\leq l \leq 50$ {\it solid red}, $2\leq l \leq 10$ {\it dashed green} and $6\leq l \leq 50$ {\it dotted blue}) and the parameter $\lambda^{PV}(r_0)$ ({\it dashed-dotted purple} line) that gives the observed bulk flow $v^{bulk}=407$ km/sec. See the magnified window for small distances  in the upper-left corner of the plot. (b): The CMB $S/N$ associated with the SSCD for $\lambda^{PV}(r_0)$. The $S/N$ is calculated for different multipole ranges (the {\it solid blue} line is for the $2\leq l \leq 50$   and  the {\it dashed red} one is for $6\leq l \leq 18$). }
\end{figure}

As we can see, the signal is smaller than the noise if the SSCD is located relatively nearby, i.e. $r_0< 300$ Mpc/h for the full signal and $r_0<500$ Mpc/h for the  signal of the safe modes. In this case the SSCD can induce the large peculiar velocity without leaving detectable imprints in the CMB.
The Shapley supercluster, located at around $140$ Mpc/h ($z\sim 0.046$) is believed to be responsible for a good fraction of the bulk velocity \cite{Smith, Bardelli}, and it is a possible candidate for a SSCD\footnote{We thank A. Dekel for mentioning this possibility to us.}. Needless to say, it should be interesting to see whether the Shapley supercluster fits the profile discussed in Section 3. If indeed there is a SSCD so close to us then it is natural to suspect that there are others in the visible universe and look for their imprints.

At larger distances, we see that $S/N >1$ and so the imprints of a SSCD with $\lambda=\lambda^{PV}(r_0)$ in the CMB should be detectable. There is, however, a dip in the $S/N$ of the safe modes at $r_0\sim 5600$ Mpc/h (see Figure 6 (b)). This dip in the signal is so extreme (it goes down all the way to $S/N\cong2$) that such a SSCD is hard to detect.

Nevertheless we would like to suggest now that perhaps there might already be hints for such imprints in the WMAP data.
As follows from the discussion in the previous sections, the CMB imprint associated with a SSCD located at $r_0\sim 4400$ Mpc/h $-~4900$ Mpc/h would be a spherically symmetric hot or cold spot localized roughly in the same direction as the bulk flow. Indeed, such a region is visible in the Kp0 foreground reduction mask\footnote{The Kp0 mask is the most aggressive of high intensity masks used to filter out "bright" sources. It is based on the K band map and includes a 0.6 degree radius exclusion area around known point sources, amounting in total to $23.2\%$ of the sky.} of WMAP (see Figure 7).
Moreover, \cite{Hajian}  argued  that this hot spherically symmetric region of  $\Delta\theta\sim16^{\circ}$ of the Kp0 mask is responsible for a good portion  of the  observed correlations at large distances (which in turn are considerably smaller than predicted by the $\Lambda$CDM model).

It is  tempting to speculate that this hot spot and anomalous peculiar velocity are induced by a single SSCD.
Indeed a SSCD with $\lambda\sim 150$ and $r_0\sim 4500$ Mpc/h can explain both anomalies. Keeping only the safe modes, $6\leq l\leq18$,  we see (figure 6 (b)) that the $S/N$ associated with such a SSCD is  $\sim 4$, which means that it is detectable, but not necessarily a prominent signature that  would pop up in regular analyses of  the CMB sky.

As shown in Figure 7, this hot spot, located at $(l,b) \sim (244^\circ- 276^\circ,\pm 13^\circ)$  in galactic coordinates, is slightly away from the direction of the bulk flow as reported in \cite{Watkins}.
This can be partially explained if we take into account the statistical fluctuations in the peculiar velocity mentioned in the last section. Using the values $v_{rms}\sim110$ km/sec and $v_{bulk}\sim407$ km/sec, we get an upper bound of  $\sim14.4^{\circ}$ on the statistical error of the direction measurement of the bulk flow velocity. Together with the reported measuring error from \cite{Watkins} this yields a crude total error estimate for the direction of the bulk flow. We should mention that in the last decade alone there have been several reports of bulk flow measurements with different results. In Figure 7 we show the recent result of \cite{Watkins}, which incorporates the results from several large scale surveys, a previous result of a combination of surveys \cite{OldWatkins} and the result of \cite{SMAC}, which is based solely on the SMAC survey and lies between the first two and agrees much better with the direction of the hot spot.

We emphasize that current data is not sufficient to reach a definite conclusion about the nature of this hot spot. It is not even  clear that a cosmological explanation is needed. In particular, it might be explained via the SZ effect associated with electrons in our galaxy \cite{SZe,Waelkens}. The nice feature of this explanation is that our location in the galaxy fixes the location of the hot spot  in the right place to agree with experiment. However, the effect appears to be too small by roughly two orders of magnitude. We  hope that near future measurements will provide a better understanding of  this hot spot and of the bulk flow.

\begin{figure}\label{fig:WMAP}
\begin{picture}(200,150)(0,0)
\vspace{0mm} \hspace{0mm} \mbox{\epsfxsize=150mm
\epsfbox{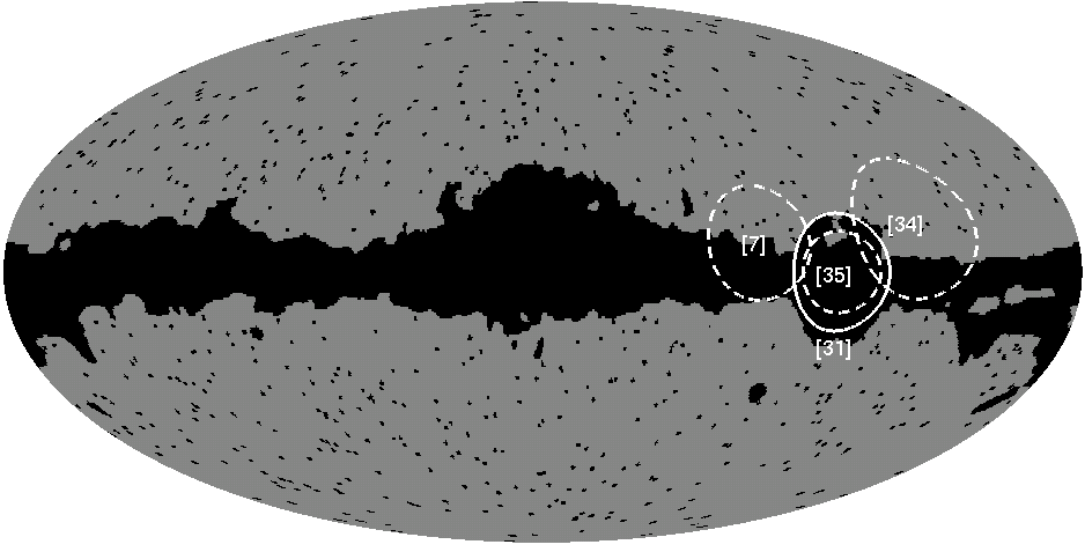}}
\end{picture}
\caption{A possible correlation between the anomalously large bulk flow and the origin of CMB correlations at large scales: we plot the Kp0 mask of WMAP and superimpose the spherically symmetric region singled out by \cite{Hajian} as a major contributor to correlations at large angular scales and the direction of the bulk flow according to three reports \cite{Watkins, OldWatkins, SMAC}, allowing a one $\sigma$ error (combination of measurement error reported by \cite{Watkins, OldWatkins, SMAC} and statistical error calculated in section 5).}
\end{figure}

\sectiono{Summary and discussion}

In this paper we studied the cosmological imprints of a pre-inflationary particle (PIP). Let us summarize our findings:

\begin{itemize}

\item A PIP provides the seed for a spherically symmetric cosmological defect (SSCD). The shape of the SSCD is fixed and its magnitude is linear in the parameter $\lambda$ which is determined by the PIP's mass (see (\ref{wq})).

\item A SSCD located within a large scale survey is detectable ($S/N >1$) if it was seeded by a PIP with $\left|\lambda\right|\gtrsim1$. This statement depends only logarithmically on the size of the large scale survey.

\item The CMB imprints associated with a SSCD located at a distance $r_0$ from us are detectable only for $\left|\lambda\right|>\lambda_{cr}(r_0)$. In the range of current and near future large scale surveys, $ r_0 \lesssim 500$ Mpc/h , we find $\lambda_{cr}(r_0)$ to be much larger than $1$. Hence it should be easier to detect a nearby SSCD via its imprint on structure formation.

\item At larger distances $\lambda_{cr}(r_0)$ drops and reaches a minimum at $r_0=1500$ Mpc (which corresponds to $z\sim 0.575 $), where $\lambda_{cr}=4.7$. At even larger distances $\lambda_{cr}(r_0)$ starts to grow because of partial cancellations between the SW and the ISW contributions.

\item Due to the SW-ISW cancellation the CMB imprint for $4200$ Mpc/h $< r_0 <4900$ Mpc/h is a {\it localized} spot in the sky. The size of the spot is highly sensitive to $r_0$ in this range.

\end{itemize}

We have also entertained the possibility that a SSCD can explain some of the potential large scale anomalies:

\begin{itemize}

\item A SSCD with $\lambda\sim-95$ located at $r_0\sim 4700$ Mpc/h creates a cold spot of the correct angular size and amplitude to fit the WMAP Cold Spot. To study the accordance better one needs to know the shape of the Cold Spot with better resolution.

\item A SSCD might explain the anomalously large bulk flow reported in \cite{Watkins,Lavaux:2008th}. A nearby SSCD, $r_0\leq300$ Mpc/h, can easily induce a large enough bulk flow velocity while having negligible CMB imprints ($S/N<1$). A SSCD with $\left|\lambda\right|\sim 170$ located at $4400$ Mpc/h $\leq r_0\leq4900$ Mpc/h yields a bulk flow of the observed magnitude and its CMB imprints are fairly small.

\item Another interesting fact is the possible connection between the location of the hot spot singled out in the galactic plane by \cite{Hajian} and the direction of the bulk flow. A single SSCD can explain this correlation neatly: If it is located
    at $r_0\sim 4500$ Mpc/h  and it has $\lambda\sim 150$ it will produce both a hot spot of roughly the same angular size as that of \cite{Hajian} and a peculiar velocity field of proper magnitude \cite{Watkins, Lavaux:2008th}.

\end{itemize}

We expect future experiments to provide a wealth of new information. In particular, wider frequency bands (as in PLANCK) should provide a better view of the galactic plane and beyond it. Therefore, more could be said about the spherical region \cite{Hajian} and its possible link to the bulk flow, as well as how good the proposed SSCD explains both phenomena. In addition, improved  large scale surveys  will provide a better bulk flow measurement by decreasing the root-mean-square peculiar velocity error. This would specify the location of the SSCD more precisely. We could also question the nature of non-gaussianities in the CMB and try to trace their origin in the bulk-flow-correlated coordinate system.

To complete the discussion, we raise the  question: in which pre-inflationary scenarios we expect to find a PIP with a large enough $\left|\lambda\right|$ to induce noticeable cosmological imprints? This is a natural question as $\lambda$ is a dimensionless parameter that is much smaller than $1$ for all standard model particles, and hence the cosmological imprints of a single standard model PIP is undetectable.

A particular pre-inflationary scenario in which such a large $\lambda$ is expected was proposed recently in \cite{IK} (for earlier and closely related work see \cite{krs}). In this scenario, all kinds of particles are  produced at the pre-inflationary period (either thermally or via a dynamical mechanism such as in \cite{Chung:1999ve,Kofman:2004yc}). Particles which satisfy $m^{'} V^{'}<0$ have two appealing features compared to those with $m^{'} V^{'}\geq 0$. First, they are extremely effective in resolving the overshoot problem associated with small field models of inflation. Second, they are light when they are produced and they become heavier as the inflaton rolls down towards the slow-roll region. For example, the mass of the particles discussed in \cite{IK} grows exponentially fast with the inflaton. Therefore, such particles are easy to create and by the time the inflaton is in the slow-roll region they can easily have a large enough $\lambda$ to seed a noticeable spherically symmetric over dense region.

Another possible scenario is the following: suppose that the pre-inflationary phase includes many PIPs, some with  $m^{'} V^{'}< 0$ and roughly the same number with
$m^{'} V^{'} > 0$ and suppose that the individual $|\lambda|$ of each of these particles is much smaller than $1$. Being a scalar field, inflaton exchange would cause particles with the same sign of $m^{'}$  to attract and particles with opposite sign to repel\footnote{We thank N. Afshordi for pointing to us  the potential importance of this interaction.}. This could lead to clumping of particles with the same sign of $m^{'}$ which effectively gives\footnote{Our current rough estimate indicates that typically the particles do not have enough time to clump. This interesting dynamical process deserves however a more detailed study.}
$
\lambda_{eff}=\sum_i \lambda_i,
$
(the sum is over particles with the same sign of $m^{'}$).
Though each $|\lambda_i|$ can be much smaller than $1$, with enough particle clumping we could have $|\lambda_{eff}|\geq 1$. Note that in this scenario, unlike in the previous one, we can get also a detectable under dense SSCD.

\vspace{10mm}

\newpage

\noindent {\bf Acknowledgements}

\vspace{4mm}

This work is supported
in part by the Israel Science Foundation (grant number 1362/08) and by the European
Research Council (grant number 203247).

\appendix
\sectiono{Derivation of  (\ref{eqmotion}) and (\ref{yuj})}

In this section we calculate the primordial potential $\Phi_0(k)$ which
serves as an initial condition for the metric perturbation in the post-inflationary epoch.
First, we derive the modified equation of motion for the inflaton in
presence of the massive point-like particle, using the spatially flat slicing gauge.
Next we evolve the perturbation through the horizon and find the primordial
potential $\Phi_0(k)$ in the Newtonian comoving gauge.

\subsection{The Spatially Flat Slicing Gauge}

Let us start with deriving the modified equation of motion for the
inflaton in the spatially flat slicing gauge. This gauge appears to
be one of the most suitable to deal with perturbation theory during
inflation, since the only approximation that has to be made is the slow-roll approximation.
The perturbated metric in this gauge reads \be ds^2 =
-(1+A)^2dt^2+2a(t)B_{,i}dx^idt +a(t)^2h_{ij}dx^idx^j~~~~
\textrm{where}~~~~h_{ij} = \delta_{ij},\ee and the perturbed
scalar field is  \be \phi(x,t) = \phi(t)+\delta\phi(x,t).\ee

First, we discuss the equation of motion for the inflaton only. In
this gauge the equation of motion (see \cite{Hwang} for derivation) reads \be
\delta\ddot\phi+3H\delta\dot\phi-\frac{1}{a(t)^2}\nabla^2\delta\phi=\left(V''
 +2\frac{\dot
H}{H}(3H-\frac{\dot H}{H}+2\frac{\ddot
\phi}{\dot\phi})\right)\delta\phi.\ee The R.H.S. of the equation is close to  zero in the slow-roll approximation, hence
\be \label{beom}
\delta\ddot\phi+3H\delta\dot\phi-\frac{1}{a(t)^2}\nabla^2\delta\phi=0.\ee
The key feature of this gauge is that the equation (\ref{beom})  does not ignore the coupling of the inflaton to the curvature perturbation as it does, for instance, in the
Newtonian conformal gauge.

We will also make use of one of the perturbed Einstein equations, namely
\be
A = \frac{\dot\phi}{2H}\delta\phi = -\frac{1}{2}\frac{V'}{V}\delta\phi,
\ee
which couples the metric perturbation to the perturbation in the inflaton.

If in addition to the inflaton field there exists a PIP, which we assume to be
a non-dynamical, massive point-like particle, the following term should be added to the action \be \label{Sp}S_{particle}=- \int
d\tau m=\int dx^0 \sqrt{-g_{00}}m, \ee where we have used $d\tau =
\sqrt{-g_{00}}dt = (1+A)dt $. To first order in perturbation,
the variation of $S_{particle}$ with respect to the inflaton is \be
\delta S_{particle} = \int d^4x\left(\frac{\p m}{\p \phi}\delta
\phi+m A\right) \delta^3(x)=\int
d^4x\sqrt{-g}\left(\lambda\frac{\delta^3(x)}{a(t)^3}\right),\ee where
\be \lambda = \frac{\p m}{\p \phi}-\frac{1}{2} \frac{ V'}{V}m. \ee This additional
term modifies the equation of motion of the inflaton so that the complete equation reads
\be \label{meom}
\delta\ddot\phi+3H\delta\dot\phi-\frac{1}{a(t)^2}\nabla^2\delta\phi
 =-\frac{\delta^3(x)}{a(t)^3} \lambda.
 \ee
The solution to this inhomogeneous equation was discussed in detail
in \cite{Sunny} and is \be \label{phi}
\delta\phi(k)=-\frac{\lambda}{k^3}\frac{H
}{{\sqrt{32\pi}}}.\ee

Perturbations generated during inflation are related to the
post-inflationary quantities via a gauge invariant variable
that is conserved on super-horizon scales \be \xi(k) \equiv
- \frac{
H}{\dot\phi}\delta\phi(k)-\Psi.\ee In the gauge under consideration, the perturbation in the space-space metric vanishes $\Psi =0 $. Therefore the conserved quantity at the end of inflation reads
\be \label{xi}  \xi(k) =  -\frac{
H}{\dot\phi}\delta\phi(k).\ee

\subsection{Primordial Potential}

After inflation ends it is convenient to switch to the Newtonian
conformal gauge, which in the absence of the anisotropic stress has
the following perturbed metric  \be ds^2 =
-(1+2\Phi)dt^2+a(t)^2(1-2\Phi)dx^idx^j\delta_{ij},\ee where $\Phi$ is the
gravitational potential. In this gauge the
primordial perturbation in the gravitational potential is  \be \Phi_0(k)
= -\frac{2}{3}\xi(k).\ee

For the particular solution (\ref{phi}), the primordial potential
reads \be  \Phi_0(k) =\frac{\lambda}{k^3}
\frac{H}{12\sqrt{\pi\epsilon}} ,\ee where the
constant \be \frac{H}{12\sqrt{\pi\epsilon}} =  \frac{1}{6\sqrt{6\pi}}\frac{V^{3/2}}{V'}=
1.984\cdot10^{-5}\ee comes  from the  COBE normalization: $\frac{V^{3/2}}{V'} = 5.169\cdot10^{-4}$.

\sectiono{Superhorizon Mode - A Consistency Check}

As was shown by the authors of  \cite{Turner} for a CDM universe and \cite{Erickcek} in the  $\Lambda$CDM case, the
observed dipole induced by any superhorizon perturbation should be very small. Here we calculate the dipole created by a SSCD located far beyond the surface of last scattering and show that it is indeed negligible.
We do this in both cases (with and without the dark energy component). In each case we expand the dipole in powers of $r_{lss}/r_0$ and calculate the linear effect only.

First, let us neglect the dark energy. In this case the dipole
 consists of the peculiar velocity term  and of the gravity term, which in the CDM case is the SW contribution only
\be D^{Observed} = D^{Doppler}+D^{Gravity}.\ee Each term reads  \be \label{SW}D^{Gravity}=S_1^{SW} =
-\lambda\frac{2\tilde C }{3}\sqrt{\frac{\pi}{3}}\frac{r_{lss}}{r_0}, \ee and \be \label{DD}
D^{Doppler}=\lambda\frac{2\tilde C}{H_0}\sqrt{\frac{\pi}{3}}\left[\frac{5}{3}-D_1(0)-\frac{2H_0z_{lss}}{3H_{lss}}\right]\frac{1}{r_0}.\ee
In (\ref{DD}) the last term in brackets is  the contribution due to the motion
of the last scattering surface, which we have approximated by \be \label{vlss}\vec v^{lss}\cdot \hat n = \frac{ v(\theta =
0)^{lss}- v(\theta=\pi)^{lss}}{2}. \ee
Keeping in mind that in a standard CDM scenario  $r_{lss} = 2/H_0$ and $D_1(0) = 1$, we have
\be
D^{Doppler}=\lambda\frac{4\tilde C}{3}\sqrt{\frac{\pi}{3}}\left[1-\frac{1}{\sqrt{1100}}\right]\frac{1}{H_0r_0}.\ee
We see that the SW contribution and the Doppler term due to the peculiar motion of an observer (the first term in brackets) cancel each other. Therefore, the leading order to the observed dipole in powers of $r_{lss}/r_0$  comes solely from the motion of the plasma at decoupling. This effect is only $\sim 10^{-2}$ of each of the cancelled components. In other words, the dipole produced by a distant SSCD is suppressed and would be barely observable.

        The same conclusion can be made for the $\Lambda$CDM background. We will show that,  in agreement with
        \cite{Erickcek},   the first order contribution to the dipole vanishes in the presence of dark energy \be D^{Observed } = D^{Doppler}+D^{Gravity} = 0.\ee
        For the background under consideration, the gravity term  consists of the SW and the ISW effects $ D^{Gravity} = S_1^{SW}+S_1^{ISW}$.  The expression for the SW term is identical to (\ref{SW}). In order to calculate the ISW contribution we should
        integrate over the  trajectory of a CMB photon from the last scattering
        surface to the observer. For a superhorizon structure, each $r'$ on the trajectory of a CMB photon satisfies  $r' \ll r_0$, which allows us to find an analytical expression for the ISW term.  Following \cite{Erickcek}, we find \be
        S_1^{ISW}=\lambda\frac{2\tilde C}{H_0}\sqrt{\frac{\pi}{3}}\left[\frac{r_{lss}H_0}{3}-\frac{5}{3}+D_1(0)+
        \frac{2H_0z_{lss}}{3H_{lss}}\right]\frac{1}{r_0}. \ee The
        expression for the Doppler term is identical to (\ref{DD}). Adding up all the contributions we obtain $D^{Observed} = 0$, i.e. in the  $\Lambda$CDM universe the linear contribution to the observed dipole due to a superhorizon SSCD vanishes.

        A consequence of the above analysis is that in order to create the  anomalous bulk flow by a single superhorizon SSCD, the value of $\lambda$ should be enormous. It is hard to imagine any physical process in charge of PIPs with such large $\lambda$.

\end{document}